\documentclass[apj]{emulateapj}

\usepackage{dcolumn}
\usepackage{amsmath,amssymb,amsthm,paralist}
\usepackage{graphicx}
\usepackage{appendix}
\usepackage{subfigure}
\usepackage{longtable}
\usepackage{url}
\usepackage{verbatim}
\usepackage{lineno}

\newcommand*{\ecapture}{$e^{-}$-capture}

\newcommand*{\maxi}{MAXI~J0556-332}
\newcommand*{\Ef}{E_{\rm F}}

\begin{document}


\title{Constraints on Bygone Nucleosynthesis of Accreting Neutron Stars}



\author{Zach~Meisel}
\email[]{meisel@ohio.edu}
\affiliation{Institute of Nuclear \& Particle Physics, Department of Physics \& Astronomy, Ohio University, Athens, Ohio 45701, USA}
\affiliation{Joint Institute for Nuclear Astrophysics -- Center for the Evolution of the Elements, www.jinaweb.org}
\author{Alex~Deibel}
\email[]{deibelal@msu.edu}
\affiliation{Department of Physics \& Astronomy, Michigan State University, East Lansing, Michigan 48824, USA}
\affiliation{Joint Institute for Nuclear Astrophysics -- Center for the Evolution of the Elements, www.jinaweb.org}


\begin{abstract}


Nuclear burning near the surface of an accreting neutron star produces ashes that, when compressed deeper by further accretion, alter the star's thermal and compositional structure. Bygone nucleosynthesis can be constrained by the impact of compressed ashes on the thermal relaxation of quiescent neutron star transients. In particular, Urca cooling nuclei pairs in nuclear burning ashes, which cool the neutron star crust via neutrino emission from $e^{-}$-capture/$\beta^{-}$-decay cycles, provide signatures of prior nuclear burning over the $\sim$century timescales it takes to accrete to the \ecapture \ depth of the strongest cooling pairs. Using crust cooling models of the accreting neutron star transient \maxi, we show that this source likely lacked Type I X-ray bursts and superbursts $\gtrsim$120~years ago. Reduced nuclear physics uncertainties in $rp$-process reaction rates and \ecapture \ weak-transition strengths for low-lying transitions will improve nucleosynthesis constraints using this technique.

\end{abstract}


\maketitle

\section{Introduction}

Accreting neutron stars are unique probes of matter above nuclear saturation density and at low enough temperatures for quantum phenomena to emerge~\citep{Scha06}. Accretion drives various nuclear burning processes near the neutron star surface that depend on the accretion rate and the composition of accreted material~\citep{Keek14}. Nuclear burning regimes include stable hydrogen burning~\citep{Scha99}, unstable hydrogen burning in Type I X-ray bursts~\citep{schatz1998,woosley2004,Pari13}, unstable carbon burning in superbursts~\citep{Stro02,schatz2003,Keek11}, and each burning regime produces a characteristic nuclear abundance ``ash" distribution. Further accretion compresses nuclear burning ashes deeper in the neutron star and drives further nuclear reaction sequences~\citep{Sato79} that replace the neutron star crust with processed ashes. As a result, the ocean and crust of an accreting neutron star have a thermal and compositional structure far different than the equilibrium state expected for an isolated neutron star composed of cold-catalyzed matter~\citep{Haen90a}. 



Models of neutron star quasi-persistent transients are especially impacted by the details of accreted ashes. These neutron stars sporadically accrete matter from their accretion disks during $\sim$month to $\sim$year long outbursts. Active accretion drives nuclear reactions that deposit heat in the neutron star's ocean and crust~\citep{Haen90b,Gupt07}, and the increase in temperature brings these layers out of thermal equilibrium with the core.
When accretion ceases and the system enters quiescence, the neutron star ocean and crust cool,  and the surface thermal emission powers an X-ray light curve~\citep{ushomirsky2001,rutledge2002}. The cooling light curve reveals successively deeper layers with time and provides clues to the thermal and compositional profile as a function of depth~\citep{Brow09,Page13,Turl15}. 

The presence of Urca cooling, neutrino cooling via $e^{-}$-capture/$\beta^{-}$-decay cycles between a pair of neutron-rich nuclides in an electron-degenerate environment~\citep{Gamo41,Tsur70}, was recently identified in the crusts of accreting neutron stars~\citep{Scha14} and in the shallower ocean \citep{Deib16}. Urca neutrino luminosities depend on the energy cost for $e^{-}$-capture, i.e. the $Q$-value, as $Q_{\rm{EC}}^5$, and the ambient temperature as $T^5$. Urca neutrino cooling significantly impacts the thermal structure of neutron star crusts by acting as a high-temperature thermostat therein~\citep{Scha14,Deib15,Deib16}. In principle, all neutron-rich nuclides with an odd number of nucleons $A$ lead to some Urca cooling~\citep{Meis15,Deib16}, but the strength of Urca cooling depends on Urca nuclide abundances in the crust $X(A)$ and rates at which $e^{-}$-capture and $\beta^{-}$-decay proceed, which are quantified by weak-transition strengths $\log(ft)$~\citep{Deib16}.

Here we demonstrate that Urca cooling in neutron star crusts reaching temperatures near $T \gtrsim 10^{9} \, \mathrm{K}$ during active accretion has an observable impact on the quiescent light curve. In Section~\ref{sec:urca_pairs}, we calculate Urca cooling neutrino luminosities using $X(A)$ from calculations of neutron star surface nucleosynthesis and $\log(ft)$ derived from experimental data. In Section~\ref{sec:crust_cooling}, we add Urca cooling nuclei to a model of the quiescent thermal relaxation of the hot neutron star transient \maxi, where the high crust temperature leads to strong Urca cooling. Furthermore, we demonstrate that Type I X-ray bursts and superbursts on the accreting neutron star \maxi\ \citep{matsumura2011,Homa14} $\gtrsim 120\, \mathrm{years}$ ago can likely be excluded due to the impact Urca cooling would have on the quiescent light curve. We highlight further experimental work that is possible at present and near-future nuclear physics facilities that will improve these constraints in Section~\ref{sec:discussion}. 


\section{Urca cooling nuclei pairs} \label{sec:urca_pairs}

When ashes of surface nuclear burning processes are buried by further accretion, they enter a degenerate electron gas whose Fermi energy $\Ef$ increases with depth. When $\Ef \approx |Q_{\rm{EC}}|$ for a given nucleus, where $Q_{\rm{EC}}$ is the electron-capture threshold energy, $e^{-}$-capture ensues that preserves the mass number $A$ but changes the proton number $Z$ to $Z-1$~\citep{Sato79}. The finite-temperature environment enables $e^{-}$-capture to proceed when $|Q_{\rm{EC}}|-k_{\rm{B}}T \lesssim \Ef \lesssim|Q_{\rm{EC}}|+k_{\rm{B}}T$, where $k_{\rm B}$ is the Boltzmann constant, and provides phase-space for the product of the $e^{-}$-capture reaction, i.e. daughter, to undergo $\beta^{-}$-decay within the same depth-window~\citep{Scha14}. For cases in which the $\beta^{-}$-decay rate of the $e^{-}$-capture daughter is significant with respect to its $e^{-}$-capture rate, a condition which is fulfilled for all odd-$A$ nuclides due to the monotonic increase of $|Q_{\rm{EC}}(A)|$ for decreasing $Z$, an $e^{-}$-capture/$\beta^{-}$-decay cycle can create an Urca pair. Pairs with significant weak-transition strengths, i.e. small $\log(ft)$, exhibit rapid cycling and consequently produce large neutrino luminosities. Layers in the crust that contain Urca cycling nuclei, denoted Urca shells, limit the heat transfer between regions above and below the Urca shell.

The neutrino luminosity for an Urca shell can be expressed as
\begin{equation}
L_{\nu} \approx L_{34}\times10^{34}{\rm{erg~s}}{}^{-1}X(A)T_{9}^{5}\left(\frac{g_{14}}{2}\right)^{-1}R_{10}^{2} \ ,
\end{equation}
as derived in \citet{Tsur70,Deib16}. Here $X(A)$ is the mass-fraction of the $e^{-}$-capture parent nucleus in the composition, $T_{9}$ is the temperature of the Urca shell in units of $10^{9} \, \mathrm{K}$, $R_{10}\equiv R/(10~\rm{km})$, where $R$ is the radius of the Urca shell from the neutron star center, and $g_{14}\equiv g/(10^{14}~\rm{cm}~\rm{s}^{-2})$, where $g = (GM/R^2)(1-2GM/Rc^2)^{-1/2}$ is the surface gravity of the neutron star. The intrinsic cooling strength of the Urca pair, $L_{34}$, is given by
\begin{equation}
L_{34}=0.87\left(\frac{10^{6}~{\rm{s}}}{ft}\right)\left(\frac{56}{A}\right)\left(\frac{Q_{\rm{EC}}}{4~{\rm{MeV}}}\right)^{5}\left(\frac{\langle F\rangle^{*}}{0.5}\right) \ ,
\end{equation}
where $\langle F\rangle^{*}\equiv\langle F\rangle^{+}\langle
F\rangle^{-}/(\langle F\rangle^{+}+\langle F\rangle^{-})$, the Coulomb factor $\langle F\rangle^{\pm}\approx2\pi\alpha
Z/|1-\exp(\mp2\pi\alpha Z)|$, and $\alpha\approx1/137$ is the
fine-structure constant.

We use values for $L_{34}$ similar to those in \citet{Deib16}, which are based on $Q_{\rm{EC}}$ calculated using atomic mass excesses from \citet{Audi12} and $\log(ft)$ based on experimental values when available; otherwise, $\log(ft)$ are obtained from Table~1 of \citet{Sing98} using the experimentally determined ground state spin-parities $J^{\pi}$ of the $e^{-}$-capture parent and daughter nuclides\footnote{Evaluated Nuclear Structure Data File (ENSDF)--a computer file of experimental nuclear structure data maintained by the National Nuclear Data Center, Brookhaven National Laboratory (www.nndc.bnl.gov)--as of 2015 December 11.}. We opt for their~\citep{Sing98} ``Centroid" minus their ``Width" for a given $\Delta J$-$\Delta\pi$, corresponding to a relatively fast, but plausible weak transition rate. This generally results in $ft$ roughly one order of magnitude smaller than assumed in \citet{Deib16}, but one order of magnitude or more larger than the quasi-random phase approximation $ft$ employed in \citet{Scha14}.

\begin{figure}[t]
\begin{center}
\includegraphics[width=1.0\columnwidth,angle=0]{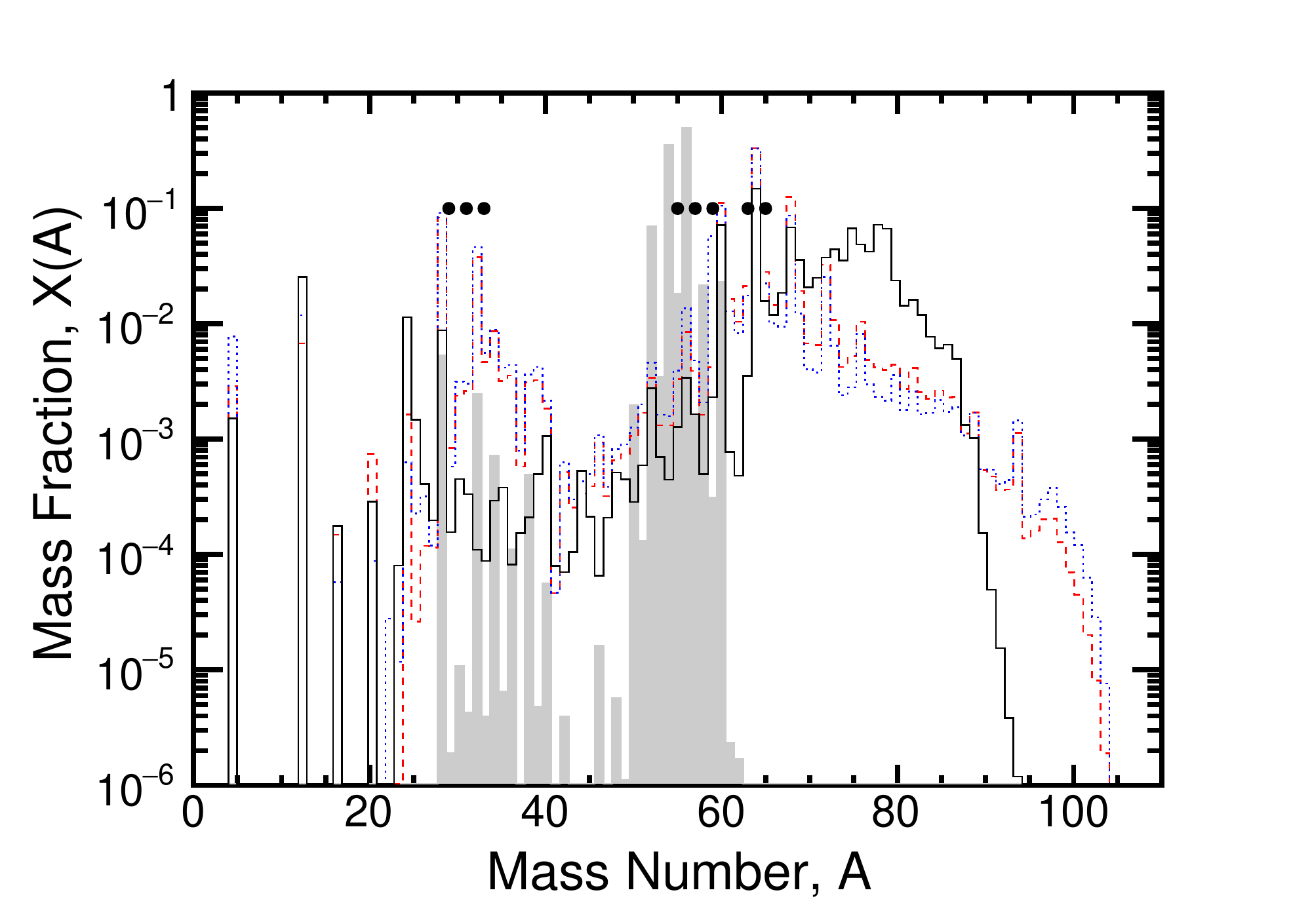}
\caption{(color online.) Abundance distributions for ashes from
model calculations of superbursts (gray-shaded histogram),
stable burning (black histogram), and Type I X-ray bursts for a
nominal nuclear reaction rate library (red-dashed histogram) and with the
$^{59}\rm{Cu}(p,\gamma)$ rate reduced by a factor of 100
(blue-dotted histogram) to demonstrate the sensitivity to nuclear
reaction rates. The black filled-circles indicate
mass-numbers with intrinsically strong Urca cooling strengths, according to \citet{Scha14}.
\label{Ashes}}
\end{center}
\end{figure}

\begin{table*}
\centering
\caption{\label{StrongPairs} Properties of the strongest Urca $e^{-}$-capture reactant nuclides, i.e. parents, identified in this work and in ~\citet{Scha14}, absent even-$A$
nuclides, excluded by~\citet{Meis15,Deib16}. Type I X-ray burst ashes from the multi-zone calculations from~\citet{Cybu16}, as well as the last burst from models 1 and 2 of~\citet{Jose10} are included for comparison.
}
\def\arraystretch{1.4}
\begin{tabular*}{0.845\textwidth}{lccrrrrrrr}%
EC Parent & $|Q_{\rm{EC}}|$~(MeV) & $\log(ft)$ & $L_{34}$ & $X_{\rm{SB}}$ & $X_{\rm{S}}$ & $X^{\rm{This Work}}_{\rm{XRB}}$ & $X^{\rm{Cyburt,MZ}}_{\rm{XRB}}$ & $X^{\rm{Jos\acute{e},1}}_{\rm{XRB}}$ & $X^{\rm{Jos\acute{e},2}}_{\rm{XRB}}$\\ \hline
$^{29}\rm{Mg}$ & 13.3 & ~5.1 & 8.2E+3 & 1.9E-6 & 1.6E-4 & 8.4E-4 & 1.8E-4 & 6.9E-4 & 6.7E-4 \\
$^{31}\rm{Al}$ & 11.8 & ~4.9 & 4.2E+3 & 4.3E-6 & 3.4E-4 & 2.6E-3 & 2.8E-3 & 1.8E-3 & 1.4E-3 \\
$^{33}\rm{Al}$ & 13.4 & ~5.2 & 3.7E+4 & 4.0E-6 & 8.8E-5 & 4.7E-3 & 5.2E-3 & 5.1E-3 & 2.4E-3 \\
$^{55}\rm{Sc}$ & 12.1 & ~4.9 & 2.4E+3 & 1.8E-2 & 1.3E-3 & 3.3E-3 & 4.7E-3 & 5.3E-3 & 4.0E-3 \\
$^{57}\rm{Cr}$ & ~8.3 & 11.6 & 8.6E-5 & 1.6E-3 & 1.7E-3 & 3.9E-3 & 4.1E-3 & 3.1E-3 & 3.4E-3 \\
$^{57}\rm{V}$ & 10.7 & ~4.9 & 1.2E+3 & 1.6E-3 & 1.7E-3 & 3.9E-3 & 4.1E-3 & 3.1E-3 & 3.4E-3 \\
$^{59}\rm{Mn}$ & ~7.6 & 11.6 & 5.2E-5 & 3.1E-4 & 2.3E-3 & 4.2E-3 & 4.5E-3 & 3.8E-3 & 3.7E-3 \\
$^{63}\rm{Cr}$ & 14.7 & 14.4 & 1.1E-6 & 6.5E-9 & 3.6E-3 & 2.1E-2 & 9.5E-3 & 3.3E-3 & 3.2E-3 \\
$^{65}\rm{Fe}$ & 10.3 & 11.6 & 2.1E-4 & 4.2E-12 & 1.6E-2 & 2.8E-2 & 1.5E-2 & 2.2E-3 & 3.1E-3 \\
$^{65}\rm{Mn}$ & 11.7 & 11.6 & 4.1E-4 & 4.2E-12 & 1.6E-2 & 2.8E-2 & 1.5E-2 & 2.2E-3 & 3.1E-3\\
\hline
\end{tabular*}
\end{table*}

We calculate the abundances of Urca pairs by modeling hydrogen and helium burning in the neutron star envelope, while relying on previous model calculations for carbon burning in the neutron star ocean. Type I X-ray burst and stable hydrogen burning
ashes were calculated using the open-source software {\tt
MESA}~\citep{Paxt11,Paxt13,Paxt15} version 8845. We primarily use
parameters employed in \citet{Paxt15} to reproduce
Type I X-ray bursts from the source GS~1826-24~\citep{Gall04,Hege07}, which produces bursts that undergo the full $rp$-process. Our model uses their~\citep{Paxt15} 305-nucleus nuclear reaction network (with rates from REACLIB V2.0~\citep{Cybu10}), an accreted composition with $X(^{1}\rm{H})=0.72$, $X(^{4}\rm{He})=0.26$, and 2\% metals with a solar composition, and an envelope base luminosity $L$=1.6$\times10^{34}$~ergs~s$^{-1}$~\citep{woosley2004}. Type I X-ray bursts were produced using an accretion rate $\dot{M}=3\times10^{-9}~\rm{M}_{\odot}~\rm{yr}^{-1}$, while $\dot{M}=4\times10^{-8}~\rm{M}_{\odot}~\rm{yr}^{-1}$ was chosen to reproduce stable hydrogen burning. $X(A)$ were extracted from the {\tt MESA} results by averaging over ash layers with converged abundances and lacking any hydrogen or helium burning, as has been done in similar studies~\citep{Cybu16}. We average over the ashes following 14 Type I X-ray bursts and for an equivalent burning time for the stable-burning ashes. $X(A)$ from \citet{Scha14} were adopted for superburst abundances and are based on models described in \citet{keek2012}. Ash abundances are shown in Figure~\ref{Ashes}. Urca cooling neutrino luminosities were then calculated from $X(A)$ results for Type I X-ray bursts ($X_{\rm{XRB}}$), superbursts ($X_{\rm{SB}}$), and stable hydrogen burning ($X_{\rm{S}}$); the 10 strongest Urca cooling pairs are listed in  Table~\ref{StrongPairs}.

We demonstrate the sensitivity of our results to variations in $rp$-process nuclear reaction rates in Figure~\ref{Ashes} by comparing to Type I X-ray burst ash abundances for a reduced $^{59}\rm{Cu}(p,\gamma)$ reaction rate, which was one of the most influential rates for odd-$A$ nuclide production identified by~\citet{Cybu16}. The sensitivity of our results to the X-ray burst model choice is demonstrated in Table~\ref{StrongPairs} by comparison to ashes calculated from other multi-zone X-ray burst studies~\citep{Jose10,Cybu16}. We note that the X-ray burst ashes from~\citet{woosley2004} are substantially qualitatively different, where odd-$A$ mass-fractions are orders of magnitude less than those found in our present calculations. Odd-$A$ nuclide abundances are also substantially lower for the single-zone X-ray burst ashes of~\citet{Cybu16}.


\section{Crust cooling with Urca pairs} \label{sec:crust_cooling}
We assess the impact of Urca cooling on the quiescent light curve of MAXI J0556-332 (``MAXI"). MAXI is one of a handful of accreting neutron star systems observed to date that have undergone an extended accretion outburst ($\sim$months) followed by a long quiescent period ($\sim$years)~\citep{matsumura2011,Corn12,Homa14}. Among accreting neutron star transients, MAXI is exceptional due to the strong shallow heat source required to match observational data of the cooling light curve~\citep{Homa14,Deib15}. The inferred strong shallow heating results in a relatively large neutron star crust temperature, meaning that Urca shell neutrino luminosities would be especially large for this object if Urca nuclides were present in the crust. Furthermore, detection of hydrogen in the MAXI atmosphere indicates hydrogen comprises a significant fraction of the accreted material~\citep{Corn12}. Therefore, MAXI is an ideal source to search for observational evidence of Urca cooling in the accreted neutron star crust.

We model the quiescent light curve of MAXI using the open-source code {\tt dStar}~\citep{Brow15,Deib15}. {\tt dStar} solves the general relativistic heat diffusion equation using the {\tt MESA} numerical library~\citep{Paxt11,Paxt13,Paxt15} for a neutron star crust using the microphysics of \citet{Brow09,Deib15}. We model the quiescent period of MAXI following a 480~day accretion outburst near the Eddington mass accretion rate $\dot{M}_{\rm Edd}\approx2\times10^{-8}~\rm{M}_{\odot}~\rm{yr}^{-1}$~\citep{Homa14}. Following the light curve fit from \citet{Deib15}, the model uses a  neutron star mass of $M=1.5~M_{\odot}$, a neutron star radius of $R=11$~km, a crust impurity of $Q_{\rm{imp}}=1$, a core temperature of $T_{\rm core} = 10^8 \, \mathrm{K}$, and a light-element envelope. In addition to heat-deposition from accretion, we include $\dot{M}$-dependent shallow heating over the pressure range $\log_{10}(P)=28.2-28.6$ (with $P$ in cgs units) inferred from the break in the MAXI light curve near $\approx 10\textrm{--}20$~days~\citep{Deib15}, where the strength of the shallow heating $Q_{\rm{sh}}$ is adjusted to reproduce the light curve at early times. In past work, it was shown that the shallow heating strength must be between $Q_{\rm{sh}} \approx 6 \textrm{--} 16 \, \mathrm{MeV}$ per accreted nucleon to match the hot surface temperature of \maxi \ at the outset of quiescence \citep{Deib15}. To investigate the impact of Urca cooling, we include an Urca shell at the $e^{-}$-capture depth for the strongest layer identified in Table~\ref{StrongPairs}, $^{33}$Al, for which $X_{\rm{XRB}}L_{34}\approx160$ at a pressure $P_{\rm{Urca}}\approx3.6\times10^{26}{\rm{erg~cm^{-3}}}\left(|Q_{\rm{EC}}|/(3.7~\rm{MeV})\right)^{4}\approx6.2\times10^{28}{\rm{~g~cm^{-1}~s^{-2}}}$~\citep{Deib16}.

\begin{figure} 
\begin{center}
\includegraphics[width=1.0\columnwidth,angle=0]{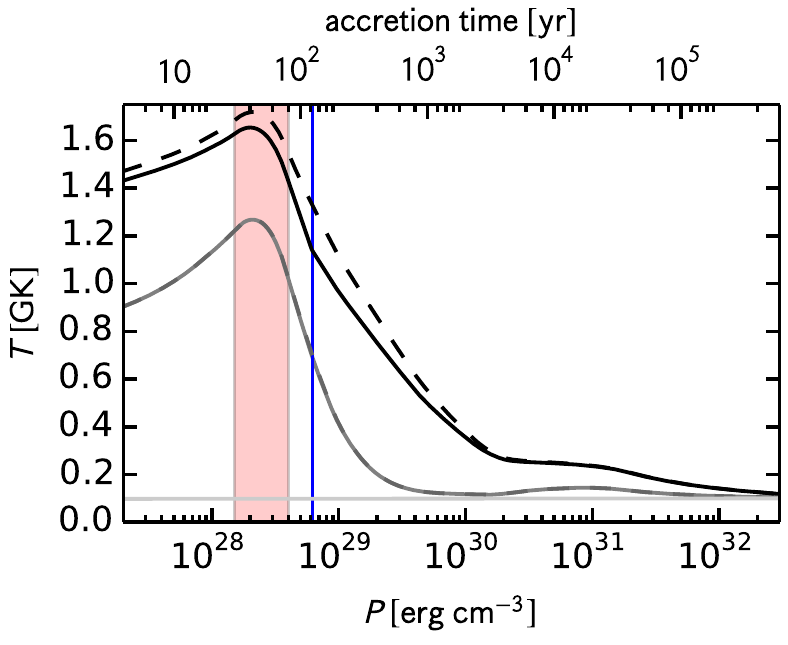}
\vspace{-0.5cm}
\caption{
 (color online.) Crust temperature profiles in MAXI, 0~days (light-gray lines), 48~days (gray lines), and 480~days (black lines) after the onset of an accretion outburst. Models with Urca cooling (solid lines) and without Urca cooling (dashed lines) are shown for the $^{33}\rm{Al}$ Urca shell assuming Type I X-ray burst ash abundances. The red-shaded area and thin blue-line indicate the zones for shallow heating and Urca cooling, respectively.
\label{Profiles}}
\end{center}
\end{figure}

Figure~\ref{Profiles} shows the temperature as a function of depth for various times during active accretion with and
without $^{33}$Al Urca shell cooling for Type I X-ray burst ashes. Shallow and deep-crustal heating drive the crust out of thermal equilibrium with the core; however, the presence of an Urca shell limits  the crust temperature by preventing shallow heating from diffusing to greater depths. This effect will impact the light curve at times $\gtrsim 20 \, \mathrm{days}$, corresponding to thermal times at depths greater than the shallow heating source.

To reproduce the observations in the absence of Urca cooling, we choose a $Q_{\rm{sh}}=6$~MeV per accreted nucleon shallow heat source \citep{Deib15}. Note that we do not fit our crust parameters to the observational data, as the purpose of this work is to demonstrate the signature of Urca cooling on formerly accreting transients. When we include the $^{33}$Al Urca shell in the crust cooling model, the predicted light curve shows a marked departure from the observations at early times. In particular, the light curve shape changes and dips below the observations between $\approx 10 \, \mathrm{days}$ and $\approx 10^3 \, \mathrm{days}$, before returning to a cooling trend that matches observations after $\gtrsim 10^3 \, \mathrm{days}$. Note that the dip caused by the Urca cooling layer alters the shape of the light curve in a way that is difficult to compensate through changes in other neutron star parameters. We have, however, attempted to counteract the effect of Urca cooling by altering other model parameters as follows.

To compensate for the Urca cooling, we first increase $Q_{\rm{sh}}$ to 8~MeV per accreted nucleon, which restores the rough reproduction of observations out to tens of days, though a marked departure is still present $\approx$100~days after the end of accretion. One might expect this signature to disappear for higher crust impurities, as higher $Q_{\rm{imp}}$ will lengthen the thermal-diffusion timescale and effectively smear-out discrete features in the crust thermal profile~\citep{Brow09,Page13}. Therefore, we adjust $Q_{\rm{imp}}$ to 100, likely the maximum plausible value~\citep{Scha01}, but find the dip in the cooling light curve near 100~days persists. We find such a signature in the light curve is present for cooling strengths $\gtrapprox1/10^{th}$ our nominal $X_{\rm{XRB}}L_{34}$ for $^{33}$Al for an accreting transient with a crust temperature $\gtrsim 10^{9}\, \mathrm{K}$. Figure~\ref{ResidLightCurves} highlights the signature of an Urca cooling layer on the transient light curve by showing the residual to our baseline {\tt dStar} calculation. Substantially stronger cooling would be required to see the impact in a transient with a crust temperature near $\sim 10^8 \, \mathrm{K}$, such as MXB 1659-29~\citep{Brow09,Deib16b}, due to the $T^{5}$-dependence of
$L_{\nu}$.

\begin{figure}
\begin{center}
\includegraphics[width=1.0\columnwidth,angle=0]{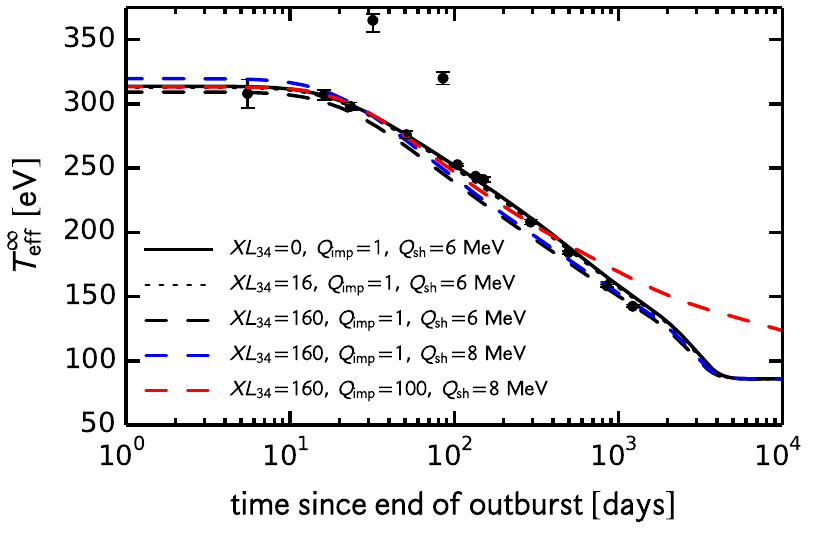}
\vspace{-0.5cm}
\caption{
 (color online.) Effective temperature of \maxi \ as a function of time for an observer at infinity. Crust cooling models are shown for various choices of the Urca cooling strength $XL_{34}$, crust impurity $Q_{\rm{imp}}$, and shallow-heating strength $Q_{\rm{sh}}$. Data points are from \citep{Homa14}, except for two points near $\sim$1000~days, which are preliminary (Aastha Parikh et al., in preparation).
\label{LightCurves}}
\end{center}
\end{figure}

\section{Discussion} \label{sec:discussion}
We have examined the impact of Urca neutrino cooling on the predicted quiescent thermal relaxation of the hot neutron star transient \maxi. Model light curves that include Urca cooling layers dip below the observations between $\approx 10 \, \mathrm{days}$ and $\approx 10^3 \, \mathrm{days}$, before returning to a cooling trend that matches observations after $\gtrsim 10^3 \, \mathrm{days}$. As a result, to fit quiescent cooling observations, we find that MAXI must be absent of Urca cooling at the strength expected for Type I X-ray burst ashes at the depth at which $^{33}$Al undergoes $e^{-}$-capture. We can also exclude Urca cooling from $^{55}$Sc $e^{-}$-capture at the strength expected for superburst ash abundances, coincidentally at nearly the same depth, since this cooling is $\approx1/3$ as strong. As such, we can constrain bygone nucleosynthesis on the surface of MAXI by calculating the time it would take for surface ashes to be buried to the depth of the $^{33}$Al Urca shell. 

For our calculations, the Urca shell depth of $P_{\rm Urca}\approx 6.2\times 10^{28}{\rm{~g~cm^{-1}~s^{-2}}}$
corresponds to a total accreted mass of $\approx4.9\times10^{27}$~g, or $\approx$120~years of constant accretion at the inferred time-averaged outburst rate $\langle \dot{M} \rangle \approx\dot{M}_{\rm Edd}\approx2\times10^{-8}~\rm{M}_{\odot}~\rm{yr}^{-1}$~\citep{Homa14}. Therefore, we conclude that MAXI likely lacked Type I X-ray bursts and superbursts $\gtrsim$120~years ago. Note that this is a lower limit that assumes MAXI has been constantly accreting at $\dot{M} = \dot{M}_{\rm Edd}$ with a $100\%$ duty cycle for the past 120 years. Our results are consistent with the near-Eddington accretion rate inferred for MAXI~\citep{Homa14,Sugi13}, which implies stable nuclear burning of accreted material on the neutron star surface~\citep{Scha99,Keek16} and that the crust of MAXI~J0556-332 is composed of stable burning ashes which have weak Urca cooling strengths. Furthermore, our results are consistent with the lack of observed Type I X-ray bursts for this source~\citep{Sugi13} and provide an additional verification of the assumed accretion rate and the assumed distance for this source.

\begin{figure}[ht]
\begin{center}
\includegraphics[width=1.0\columnwidth,angle=0]{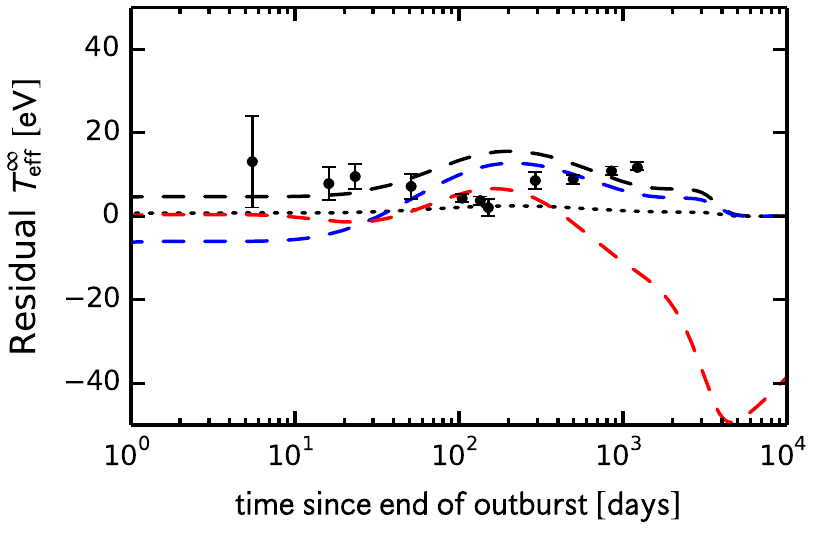}
\vspace{-0.5cm}
\caption{(color online.) Residuals for crust cooling models with Urca cooling in Figure~\ref{LightCurves} relative to the model calculation without Urca cooling assuming $XL_{34}=0$, $Q_{\rm{imp}}=1$, and $Q_{\rm{sh}}=6$~MeV.
\label{ResidLightCurves}}
\end{center}
\end{figure}

Surface nuclear burning at other timescales, such as Type I X-ray bursts at other accretion rates or different compositions of accreted material, could possibly be constrained using the technique presented here.
The \ecapture \ depths of the Urca pairs in Table~\ref{StrongPairs} span accreted masses of $\sim 10^{26} \textrm{--} 10^{28}$~g, which correspond to surface nuclear burning $\sim 2.5 \textrm{--} 250$~years ago for a neutron star with a $100\%$ duty cycle accreting at $\dot{M} \sim \dot{M}_{\rm Edd}$ and $\sim 25 \textrm{--} 2500 \, \mathrm{years}$ for accretion rates $\dot{M}\sim 0.1 ~ \dot{M}_{\rm Edd}$ typical of Type I X-ray bursters~\citep{Gall08}.
Though many of the cooling strengths ($XL_{34}$ in Table~\ref{StrongPairs}) are weak, future nuclear physics measurements and model calculations of surface nuclear burning may find stronger cooling. For varied accretion rates and nuclear reaction rates, $X$ could be enhanced by several orders of magnitude, with the few-percent level as the empirical limit for odd-$A$~\citep{Scha99,Scha01,Pari13,Cybu16}. In addition, $ft$ could be smaller by one or two orders of magnitude based on
systematics~\citep{Sing98}. Future nuclear physics measurements of particular interest are $rp$-process reaction rates impacting the production of $A=29$, 31, 33, 55, 57, 59, 63, and 65, as well as $\log(ft)$ for low-lying transitions involved in $e^{-}$-capture on $^{31}$Al, $^{33}$Al, $^{55}$Sc, $^{57}$Cr, $^{57}$V, $^{59}$Mn, $^{63}$Cr, $^{65}$Fe, and $^{65}$Mn. Such studies are possible via indirect measurements at present stable and radioactive ion beam facilities and direct measurements at near-future radioactive ion beam facilities.

The Urca cooling signature identified here in the light curves of cooling transients can be verified in the coming decades with continued monitoring of the X-ray sky by present telescopes such as MAXI\footnote{{h}ttp://maxi.riken.jp}, NUSTAR\footnote{{h}ttp://www.nustar.caltech.edu}, and ASTROSAT\footnote{{h}ttp://isro.gov/in/spacecraft/astrosat}, near-future telescopes such as NICER\footnote{{h}ttps://www.nasa.gov/nicer}, and planned telescopes such as LOFT\footnote{{h}ttp://sci.esa.int/loft}. In particular, identification of additional hot transients and long-term burst/burning monitoring for these sources would be most desirable.

In general, we find that Urca cooling in the crust has an observable impact on the light curves of transiently accreting neutron stars in quiescence whose crusts have achieved temperatures $\sim 10^9 \, \mathrm{K}$ for Urca nuclides with $X(A)\gtrsim 0.5\%$ and $\log(ft)\lesssim$5. In particular, using $\log(ft)$ derived from experimental data and data-based systematics and $X(A)$ from model calculations of neutron star surface burning conditions, we exclude the existence of Type I X-ray bursts and superbursts $\gtrsim$120~years ago for the source MAXI J0556-332. Modeling light curves of this and other neutron star transients with hot crusts after accretion turn-off can provide a window to examine the stability of surface nuclear burning on accreting neutron stars over millennia, improving constraints on the structure of accreted neutron star crusts.

\begin{acknowledgments}
We thank Laurens~Keek for providing superburst ash abundances and Jeroen Homan for sharing the last two data points of the cooling curve of MAXI J0556-332 ahead of publication (Aastha Parikh et al., in prepartion).
Z.M. is supported by the Department of Energy under grant No. DE-FG02-88ER40387. A.D. is supported by the National Science Foundation under grant No. AST-1516969. We thank the International Space Science Institute in Bern, Switzerland for support received as a part of the International Team on Nuclear Reactions in Superdense Matter. We also thank the Department of Energy's Institute for Nuclear Theory at the University of Washington for partial support during the completion of this work. This material is based on work supported by the National Science Foundation under grant No. PHY-1430152 (Joint Institute for Nuclear Astrophysics--Center for the Evolution of the Elements).
\end{acknowledgments}

\bibliographystyle{apj}
\bibliography{UrcaReferences}

\end{document}